\documentclass[preprint2]{aastex}

\shorttitle{AQUARIUS SUPERCLUSTERS. II.}
\shortauthors{CARETTA ET AL.}

\begin{document}


\title{The Aquarius Superclusters\\ II. Spectroscopic and 
Photometric Data\altaffilmark{1}}
\author{C\'esar A. Caretta\altaffilmark{2,3}, 
Marcio A. G. Maia\altaffilmark{4,2},
Christopher N. A. Willmer\altaffilmark{5,2}
}

\email{caretta@lac.inpe.br, maia@ov.ufrj.br, cnaw@ucolick.org}

\altaffiltext{1}{Partly based on observations at 
European Southern Observatory (ESO) at the ESO Schmidt, and 
the 1.52m telescope under the ESO-ON agreement; 
Observat\'orio do Pico dos Dias, operated by the 
Laborat\'orio Nacional de Astrof\'\i sica (LNA); and Complejo Astronomico El 
Leoncito (CASLEO), operated under agreement between the Consejo Nacional de 
Investigaciones Cient\'\i ficas de la Rep\'ublica Argentina and the National 
Universities of La Plata, C\'ordoba and San Juan.}

\altaffiltext{2}{Observat\'orio Nacional (ON/MCT), Rua Gal. Jos\'e Cristino,~77
-- 20921-400, Rio de Janeiro (RJ), Brazil.}

\altaffiltext{3}{Depto. de Educa\c c\~ao, Museu de Astronomia e 
Ci\^encias Afins (MAst/MCT), Rua Gal. Bruce,~586 -- 20921-030, 
Rio de Janeiro~(RJ), Brazil.}

\altaffiltext{4}{Depto. de Astronomia, Observat\'orio do Valongo (OV/UFRJ), 
Ladeira Pedro Ant\^onio,~43 -- 20080-090, Rio de Janeiro~(RJ), Brazil.}

\altaffiltext{5}{UCO/Lick Observatory, University of California, 
Santa Cruz, CA 95064, USA.}


\begin{abstract}
We present spectroscopic and photometric data for 920
galaxies selected in 68 fields of the Aquarius Cluster Catalog. 
Typically the 15 brightest candidate members with
magnitudes in the range 16 $< b_J <$ 21 were selected for
observations, and $\sim$ 71\% turn out to be cluster
members. Using the new 
redshift determinations we assign galaxies to groups and clusters,
and by including data from the literature we calculate systemic velocities and
velocity dispersions for 74 clusters, each with redshifts measured for
at least 6 individual galaxies.
\end{abstract}


\section{Introduction}

Because of the large number of galaxies contained by superclusters,
before the era of large-scale multi-object 
surveys like 2dFGRS \citep{2dF} and SDSS \citep{sdss}
much of what is known of the properties of these structures
has relied mainly on pointed observations, where only a sub-sample of
supercluster members is observed spectroscopically. Examples of
superclusters that have been studied using this strategy are
Pisces-Perseus \citep[e.g.,][]{Weg93}, 
Hydra-Centaurus \citep[e.g.,][]{daC86,daC87},
Coma-A1367 \citep[e.g.,][]{Gre78}, 
Hercules \citep[e.g.,][]{Bam98},
Shapley \citep[e.g.,][]{Qui00} and 
Corona Borealis \citep[e.g.,][]{PGH88}. 

In \citet{Car02}, hereafter Paper I,  we presented a list of 
candidates clusters and groups in the Aquarius 
region, comprising the area of sky limited 
by $22^h 57\fm0 < \alpha_{2000} < 23^h 38\fm6$, and
$-25\arcdeg 54\arcmin < \delta_{2000} < -19\arcdeg 29\arcmin$.
Of the 102 candidates, 39 were new detections.
We also presented mean redshifts for 31 previously unobserved 
clusters and improved redshifts for a further 35. 
About half of the observed candidates are 
single concentrations of galaxies,
while the remainder are superposition of two or more
poor clusters or groups along the line of sight.
Using a percolation analysis on the data we found two 
superclusters of galaxies in Aquarius, one located at $z \sim$ 0.086,
and another at $z \sim$ 0.112.
These results differ from \citet{Bat99} who
claimed finding a single 
structure of $\sim$ 110 h$^{-1}$ Mpc oriented along the line of sight.
More recent work by \citet{Smi04}, aimed at measuring the
redshift distribution of weak lensing sources, probed a part of this region
of sky using the Two Degree Field system. They detect 48 clusters in
the region, of which 22 are in common with paper I sources.

In this paper we present the spectroscopic and photometric data of our survey. 
In section 2 we describe the photometric catalogs, the target selection and
the spectroscopic procedure, from observations to redshift estimation.
The spectroscopic and photometric data for the galaxies are described in 
section 3, while the catalog of mean radial velocities and velocity dispersions 
for the clusters is presented in section 4.

\section{The Data}

\subsection{Photometric Catalog}

The selection of galaxies for spectroscopic observations is
derived from two sources:
the COSMOS/UKST Southern Sky Catalog \citep[SSC,][]{Yen92}, 
for astronomical coordinates, $b_J$
magnitudes and other photometric and shape parameters; 
and a catalog created from the digitization using the APM 
machine of $R$ band films obtained with 
1.0m Schmidt telescope of ESO (see also Paper I).
The catalog of spectroscopic candidates was limited to $b_J \leq$
20.2, where the completeness is about 85\% \citep{Car00}.
This magnitude limit is roughly the faintest that can be attained
within reasonable integration times with 2 m class telescopes, 
and is faint enough that the luminosity function is probed to 
$\sim M_{b_J}^*$  at $z \sim$ 0.2.
More than 90\% of the galaxies in the SSC have counterparts in the $R$ 
catalog (20,687 galaxies), and we estimate that the final completeness of our 
photometric catalog is above 75\% at $b_J$ = 20.2. The $b_J$ and $R$ magnitudes are
both uncertain to 0.2 mag, while the SSC positions  used in this work
are accurate to better than 1 arcsec.

\subsection{Target selection}

Since our goal was to detect the presence of superclusters of galaxies
in Aquarius, the observational strategy was designed to select
preferentially targets with a greater likelihood of being members of
clusters or groups belonging to the superclusters. This was done by
identifying significant overdensities in the projected distribution of
galaxies. In the later runs, once the 
$R$-band data became available, an additional selection criterion was
used, which took into account
the proportion of red galaxies, under the assumption that most of
these would be early-types. This means that our redshift catalog has
not sampled the region uniformly and is biased towards red galaxies.
 
As described in Paper I, the galaxies were selected from a catalog
of 102 potential clusters or groups of galaxies. 
For 68 of these, we selected galaxies located in square fields 
of $10'$ centered on the surface density peaks within the 
Abell radius. 
These typically contained all galaxies in the clusters core, 
with additional objects that could be part of lower density 
regions of the supercluster.

A total of 94 square fields were observed, and in cases where previous
measurements existed in the literature, our observations would
supplement the existing redshifts so that at least 10 galaxies would
be available to provide a robust velocity dispersion estimate.
The average surface density of galaxies for the entire Aquarius region
is $\sim$~0.1~gal~arcmin$^{-2}$ to $b_J =$ 20.2; the mean surface
density of the fields selected for observations ranged from
0.13 to 0.58~gal~arcmin$^{-2}$. 
The galaxies selected for observations span from $b_J =$ 16.0 to 21.0,
usually among the 15 brightest galaxies in the field. Obvious
foreground galaxies were excluded, so that typically 10
galaxies were observed per $10'\times10'$ field, which implies in
a sampling rate of $\sim$ 34\%.

Whenever possible, the slit was rotated so that two or more galaxies
were observed simultaneously. As a result of this strategy, there are
27 additional serendipitous galaxies fainter than $b_J =$ 20.2 in our
catalog. 

\subsection{Observational Procedure}

We carried out 25 observing runs between 1994 and 2000, using three telescopes:
the 1.52m of La Silla Observatory (ESO, Chile); the 1.60m of
Observat\'orio do Pico dos Dias (OPD/LNA, Brazil)  and the 2.15m of  
Complejo Astronomico El Leoncito (CASLEO, Argentina). 
In both 1.52m and 1.60m 
telescopes we used Boller \& Chivens spectrographs, while the 2.15m
telescope had a REOSC spectrograph. All spectrographs were mounted at the
Cassegrain focus. The log of the observing runs is shown in
Table 1, which also presents the instrumental configuration.
In Table 2 we present the details about the detector characteristics 
for the three used instrumental set ups.

For all observations, we covered the wavelength range ($\sim$
4000-7700 \AA), using 300 l/mm diffraction 
gratings for 1k pixels CCDs and 600 l/mm gratings for the 2k detectors.
This range was chosen in order to enclose most of the main 
absorption and emission optical lines for galaxies from $z$ = 0.0 to 0.2.
It contains, in absorption, the Ca{\sc ii} K and H lines, the G band, 
H$\beta$, Mg{\sc i}, Ca+Fe and Na{\sc i} D band. 
The emission lines include [O{\sc ii}], H$\beta$, [O{\sc iii}] (at 4958.9 and
5006.8 \AA), [O{\sc i}], [N{\sc ii}] (6548.1 and 6583.6 \AA), H$\alpha$ and 
[S{\sc ii}] (6716.4 and 6730.8 \AA).

Integration times ranged from 20 to 120 min, depending on the magnitude 
and surface brightness of the objects.
Longer exposures were divided in multiples of 20 min for an efficient
removal of cosmic rays events. Line comparison lamps
(He-Ar, He-Ar+Ne, He-Ar+Fe or Fe-Ar) were 
obtained immediately before or after each exposure.
Sequences of bias, domeflats and dark exposures were obtained every night,
and also some twilight sky flat-fields for each run.
Observations were done typically in dark and grey time.

\subsection{Data Reduction}

The data reduction followed a standard procedure using 
{\sc iraf}\footnote{The Image Reduction and Analysis Facility (IRAF)
is distributed by the National Optical Astronomy Observatories.}. 
This involved removal of bias; subtraction of dark current 
(only necessary in the first few observing runs); 
division by flat-field and, in a
few rare cases, correcting for illumination using twilight flats;
removal of cosmic rays; extraction of the 1-D spectrum; and
sky-subtraction. 
In the process of wavelength calibration and linearization, 
we used arc lamp spectra extracted
using the same apertures as the object frames. 
The wavelength
solutions used seventh- or eighth-order legendre polynomials and
the final fits typically used more than 40 lines with rms 0.4\AA,
0.15\AA \ and 0.09\AA, respectively for CASLEO, OPD and ESO.
In the cases of galaxies for which multiple exposures were taken,
these were co-added after linearizing.


\subsection{Redshift Determination}

Redshifts were measured using the cross-correlation 
\citep{Ton79} and emission lines analyses, both available
in the {\sc rvsao} package \citep{Kur98}. 
In the cross-correlation we used a set of 15
high-signal to noise  template spectra, most derived from  spectra
in the Southern Sky Redshift Survey \citep[SSRS,][]{daC98} database. 
The position, height and width of the highest peak for each template was
obtained by fitting a parabola. The significance value of this peak is
estimated comparing its height relative to typical noise peaks, 
and is measured by the ``R'' parameter.
In general, most of the templates give estimates within a velocity
interval of about 200 km s$^{-1}$, and for the final cross-correlation velocity
we took the one with the highest signal ``R'' value. 

\begin{figure*}[t]
   \includegraphics[scale=0.8]{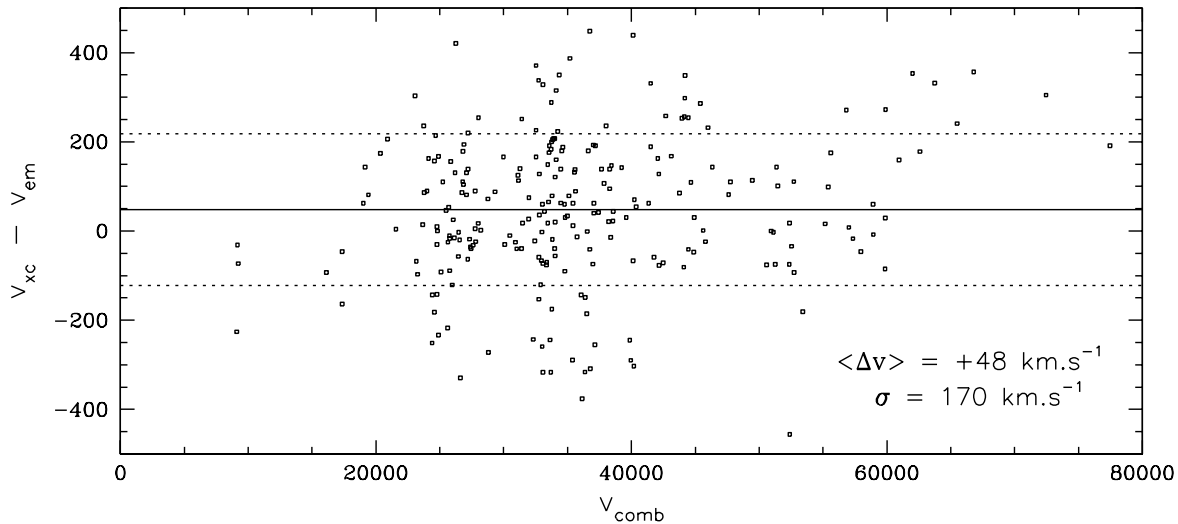}
\vspace{-8 cm}
   \caption{Residuals of cross-correlation minus emission-lines velocities 
as a function of combined velocity. Solid line represents the mean
residual value, while dashed lines represent the dispersion.}
   \label{fig1}
\end{figure*}

Whenever emission-lines are present, they are fit by Gaussians, and a
final emission velocity is obtained by weighting the redshift measured
for each individual line by its error. Some spectra have both
cross-correlation and emission-line redshifts.
In these cases,
for a difference between absorption and emission velocities less than
500 km/s, we considered a combined velocity (mean weighted by the
respective errors). The average absorption-emission difference 
was +48 km s$^{-1}$, with a rms of 170 km s$^{-1}$, 
as shown in Figure \ref{fig1}.

\begin{figure}[h]
   \includegraphics[scale=0.35]{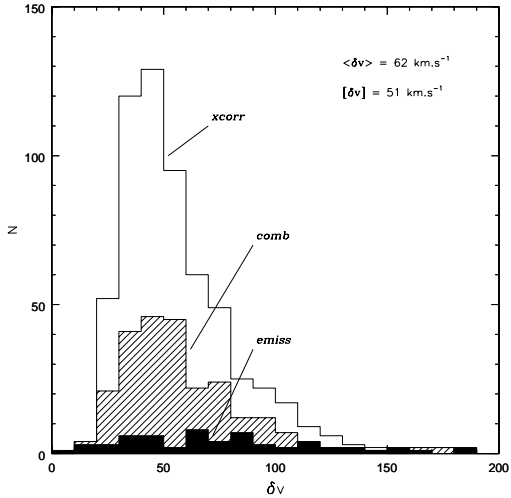}
   \caption{Distribution of velocity errors according to the different types of spectra.
$<\delta v>$ and [$\delta v$] refer to, respectively, average and median values for all data.}
   \label{fig2}
\end{figure}

\begin{figure}[h]
   \includegraphics[scale=0.35]{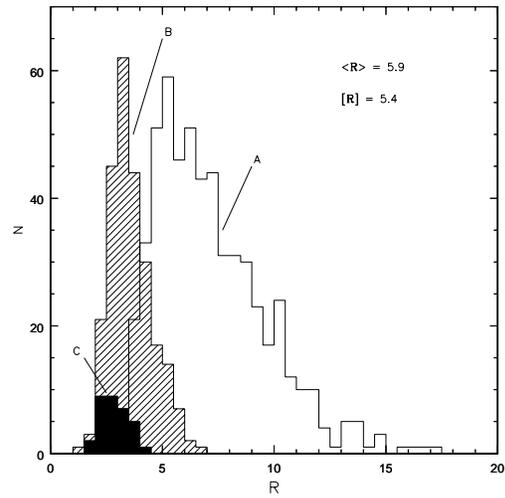}
   \caption{Distribution of ``R'' parameter for quality classes A, B and C.
$<$R$>$ and [R] refer to, respectively, average and median values for all data.
Note that the quality class is clearly related to the ``R'' parameter.}
   \label{fig3}
\end{figure}

The {\sc rvsao} package also estimates internal uncertainties 
($\delta$$v$) for the redshift measurements. 
In the case of cross-correlation redshifts,
these are related to the ``R'' parameter, while for emission-line
redshifts, the error is estimated by a combination of the line
centering error with the dispersion between the measurements of the
different lines. For galaxies with both cross-correlation and
emission-line redshifts the final internal errors are calculated in
quadrature. The distribution, average and median values of these errors
and of the ``R'' parameter are shown in Figures \ref{fig2} and 
\ref{fig3}.

All spectra were visually inpected in order to validate the redshift
and also to assign a redshift quality. 
The criteria for defining these quality flags are
summarized below: 
 
\begin{itemize}
\item{[A]} excellent, absorption and/or emission lines are well
  defined and the ``R'' parameter (signal-to-noise ratio of the
  correlation peak) much greater than 3; 

\item{[B]} good, absorption and/or emission lines confirmed by visual
  inspection or similar radial velocities for most of the templates 
  but small ``R'' (between 2 and 4);

\item{[C]} marginal, ``R'' greater than 2, but with features
  either in absorption and/or emission too weak to be 
  firmely confirmed;

\item{[D]} failed redshift, ``R'' $<$ 3 without any
  identifiable spectral feature.
\end{itemize}

\begin{figure}
\hspace{-0.5cm} \includegraphics[scale=0.42]{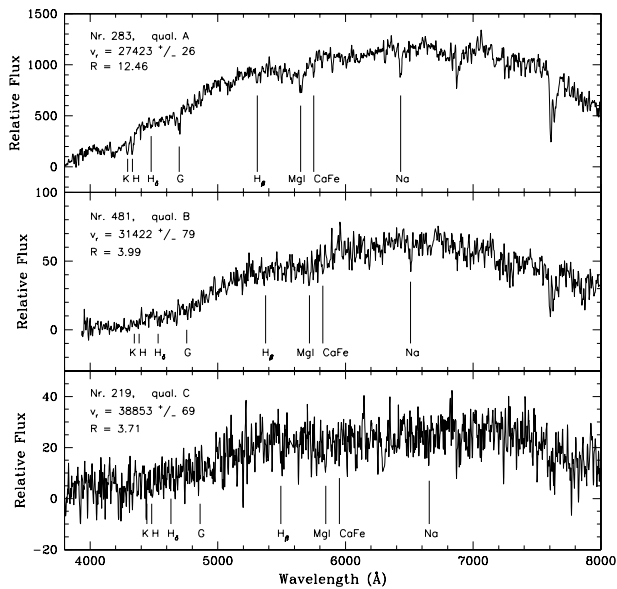}
\vspace{-0.5cm}
   \caption{Sample of absorption-lines spectra of the three quality classes.
Objects are identified by their number in Table 5 (Nr.). Also are displayed their 
radial velocity (v$_r$) and ``R'' parameter of cross-correlation.}
   \label{fig4}
\end{figure}

\begin{figure}
\hspace{-0.5cm} \includegraphics[scale=0.42]{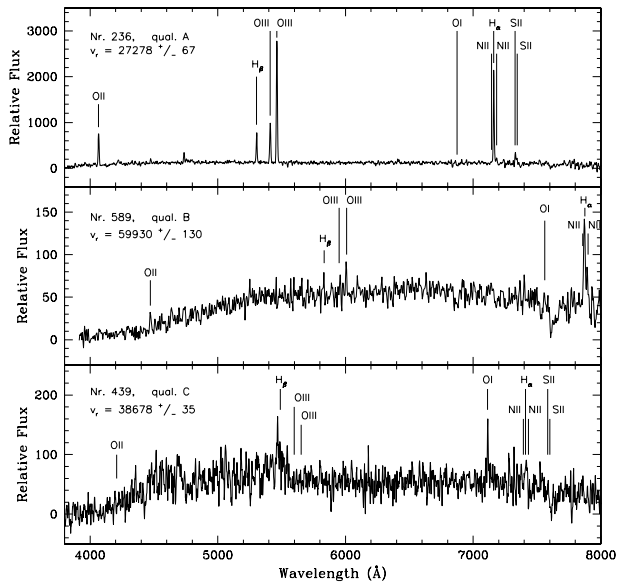}
\vspace{-0.5cm}
   \caption{Sample of emission-lines spectra of the three quality classes. Identification
of objects is the same as in Figure 1.}
   \label{fig5}
\end{figure}

\begin{figure}[t]
   \includegraphics[scale=0.42]{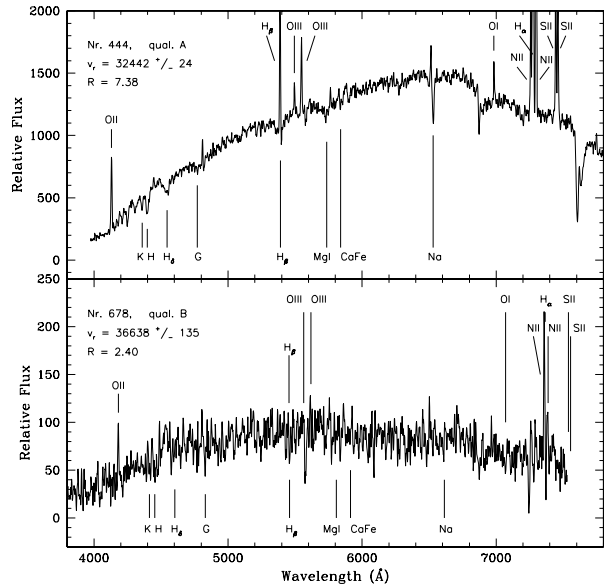}
   \caption{Sample of combined absorption and emission spectra, same identifications as in
Figure 1.}
   \label{fig6}
\end{figure}

\begin{figure}
\hspace{-1.5 cm} \includegraphics[scale=0.5]{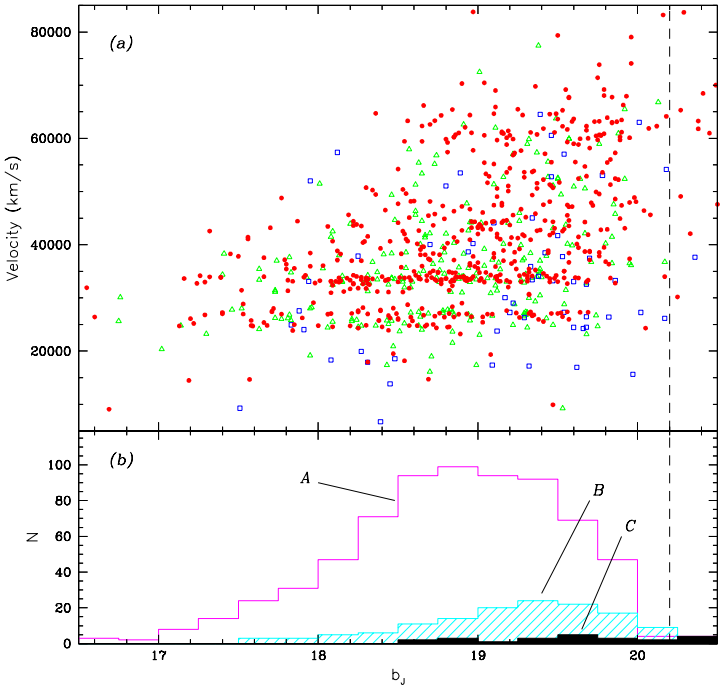}
   \caption{(a) Distribution of measured velocities with magnitudes. 
Solid circles are absorption-line
only spectra and open symbols are spectra with emission-lines 
(triangles for both emission and absorption spectra and squares for 
only emission ones). Note that absorption line spectra galaxies 
(probably early-type ones) are slightly more concentrated than the 
emission line galaxies (probably intermediate and late-type ones).
(b) Distribution of quality classes with magnitude.
The vertical dashed line represents the $b_J$=20.2 magnitude limit of
the sample.}
   \label{fig7}
\end{figure}

The fractions of spectra in each of these classes are 74\%, 15\%,
3\% and 8\%, respectively for A, B, C and D.
In all subsequent analyses only galaxies
with redshift qualities of ``C'' or better will be considered.
Typical examples for each of these redshift qualities are shown in
Figure \ref{fig4}, for spectra with only absorption-lines; Figure \ref{fig5}, for
galaxies with emission lines only; and Figure \ref{fig6}, for galaxies with 
both emission and absorption. 
Figure \ref{fig7}a shows the c$z$-magnitude diagram where solid circles
represent galaxies with cross-correlation velocities only, open triangles galaxies
with emission-line velocities only and open squares galaxies with both.
The distribution of galaxy velocities in Aquarius clearly shows two 
concentrations, one at  c$z \sim$~26,000 and another at c$z \sim$~33,500,
corresponding to the superclusters described in Paper I. 
The bottom panel of Figure \ref{fig7} shows the distribution of redshift
quality as a function of magnitude. It is clear that as fainter
magnitudes are being probed the number of quality ``A'' redfshifts
(mainly due to cross-correlation) decreases dramatically.

\subsection{Redshift Uncertainties}

A first check of the reliability of the bulk of our data is done by
separating it in sets coming from the three distinct telescopes: 
C215, O160 and E152 (see tables 1 and 2).
Internal velocity errors for all the observed galaxies in each telescope
are presented on Table 3, which reveals that ESO data
are slightly better (smaller errors), since higher resolution gratings 
(600 l/mm) were used in this set up. CASLEO data, on the other hand, 
have slightly higher internal errors, probably due to the noisest CCD
used.

Another internal evaluation of redshift uncertainties is done 
using the radial velocity standard galaxies, observed at least once
in every observing night. These are listed on Table 4.
Considering only the runs with more than 4 observations of such 
objects we could estimate the stability in the velocities measured
from each telescope. 
For ESO we have six runs (with 6 to 12 observations each), 
that give a mean dispersion of 25 km s$^{-1}$. 
For OPD we have also six runs (with 4 to 6 observations 
each) and 30 km s$^{-1}$. 
For CASLEO we have only one object in one run, and 28 km s$^{-1}$. 
This shows that all the three instrumental set ups 
produce similar internal dispersions, of about 25-30 km s$^{-1}$,
for the radial velocity standards. Since they are the best signal
to noise data, their dispersion may be considered as a lower limit
for our internal uncertainties.

We also used the standard galaxy measured velocities for external
comparison, in order to estimate possible zero points between the 
instrumental set ups used. Column 6 of Table 4 lists the
mean velocity differences between ESO and NED velocities, which give, 
on mean, +15$\pm$38 km s$^{-1}$ (for 6 galaxies). 
Similarly, column 11 show these differences between OPD and NED 
velocities (also for 6 galaxies), +45$\pm$70 km s$^{-1}$ on mean, 
and column 16 for CASLEO and NED (for only one galaxy), $-$31 km s$^{-1}$.
The same results are found using mean NED velocities (column 23).
Nevertheless, it may be noted that NED velocities are not necessarly the 
best ones for comparison since they come from different
sources ($\Delta$$v$ is not systematic for a single instrumental set up).


\begin{figure*}
\hspace{1cm} \includegraphics[scale=0.7]{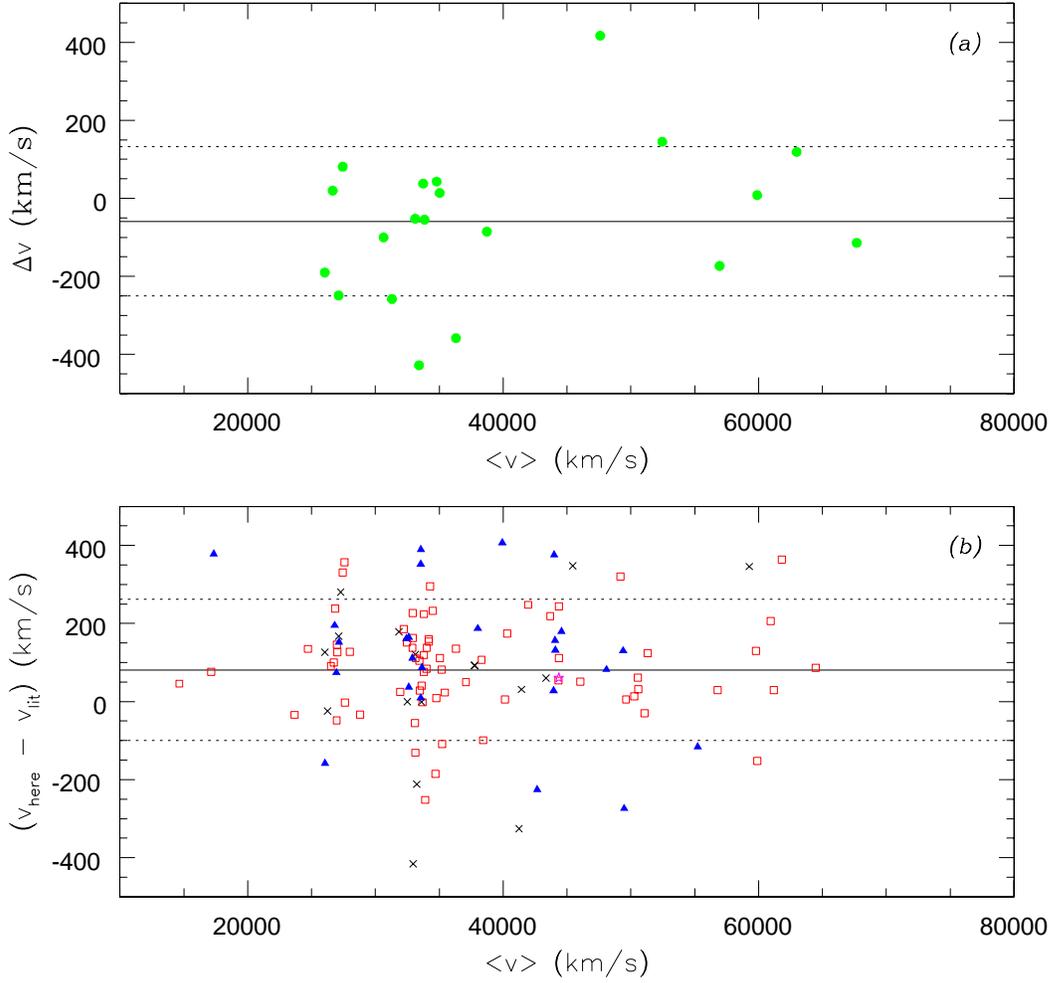}
   \caption{(a) Velocity residuals from our repeated observations
(solid circles). 
(b) Velocity residuals between our data and data from the literature,
separated in datasets from distinct telescope: open squares = E152 $\times$ 2dFGRS;
solid triangles = O160 $\times$ 2dFGRS; open star = C215 $\times$ 2dFGRS; 
crosses = all telescopes $\times$ other sources from the literature.}
   \label{fig8}
\end{figure*}

A further estimate of the redshift uncertainties
can be made using galaxies that have more than one measurement 
carried out by
ourselves, as well as comparing our measurements with those available
in the literature. 
For a total of 20 galaxies we obtained 
spectra in different observing runs. From these measurements, shown
in Figure \ref{fig8}a, we find a dispersion of 191 km s$^{-1}$.
Since the number of objects is very small, the dispersion may
be overestimated.
In Figure \ref{fig8}b we show the comparison for 118 galaxies 
of our database with redshifts also available on the 
literature. In this case, we find a 
zero-point shift of +81 km s$^{-1}$ and standard deviation of 
181 km s$^{-1}$. 
Most of these galaxies were observed as part of the 2dFGRS \citep{2dF},
represented by the symbols other than crosses. 
When the 97 2dFGRS galaxies are considered the
mean difference is +91$\pm$156 km s$^{-1}$.
The best estimation of spread in 2dFGRS itself, as measured by the authors, 
comes from their comparison with Las Campanas Redshift Survey
\citep[LCRS,][]{She96} data, with gives 109 km s$^{-1}$.
Considering equally distributed errors for the two sources, the 2dFGRS
error may be about 77 km s$^{-1}$. 
By subtracting quadratically this error to the above measured dispersion, 
this gives an estimation of the overall error of our data of 136 km s$^{-1}$.
If we now separate the data by instrumental set up (indicated by 
the different symbols in Figure \ref{fig8}), we find a mean difference
of +93$\pm$226 km s$^{-1}$ for OPD data (26 galaxies) and 
+90$\pm$123 km s$^{-1}$ for ESO data (70 galaxies). For CASLEO the 
estimation is not possible due to the small number of objects.
Thus, both OPD and ESO data have the same zero point related to 2dFGRS
data, but the dispersion of OPD data is somewhat higher than ESO data
one. Note that the number of galaxies in OPD data is about one third 
of ESO one, what may be effecting in the large dispersion measured.
Since ESO data represent 63\% of our overall sample, OPD 34\%, and CASLEO only 3\%,
we can estimate the typical uncertainty in our measured radial velocities
by the the weighted mean of ESO and OPD dispersions, 150 km s$^{-1}$,
which is consistent with all the results above.

\section{The Aquarius Galaxy Catalog}

A total of 920 new galaxy redshifts, with quality grades ``C'' or better,
were measured in the direction of
72 cluster and group candidates, 68 of which within the bounds of the
region considered in this work. In addition, we find
2851 redshifts in the literature for the same area of sky, most (2378)
coming from the 2dFGRS, but which only contains galaxies located on
the southern part of Aquarius region ($\delta < -23.5$).

Our new measurements are listed in Table 5, where we show, in
column (1), a sequential number for galaxy identification; in
columns (2) and (3), the equatorial coordinates (J2000.0); and
b$_J$ and R magnitudes in columns (4) and (5), respectively.
When a galaxy is assigned to a group or cluster, column (6) gives
the name of the cluster/group; whenever these are assigned to
different systems along the same line of sight, suffixes A, B, C are
appended to the group name; suffixes 1 and 2 are assigned to the
clusters A2538 and A3985, which are only marginally separated 
in the redshift space.
Columns (7) and (8) list the heliocentric
velocity and velocity error of the galaxy.
The codes on column (9) indicate the type of redshift 
(x = cross-correlation, e = emission lines, and c = combined 
cross-correlation and emission lines redshift), the quality grade
and the value of the ``R'' parameter. 
The observing run, as listed on Table 1, during which the galaxy spectrum 
was taken, is shown on column (10).

\section{Cluster Systemic Velocities and Velocity Dispersions}

Combining the data from 
the Aquarius Galaxy Catalog with the 2851 redshifts from the
literature we re-determine the redshifts and velocity dispersions for
several of the clusters and groups within the surveyed region.
Currently there are 81 candidates of the Aquarius Cluster Catalog
with redshifts for galaxies inside their estimated Abell radius, 9
more than published in Paper I.
As described in Paper I, about half of the observed candidates show 
more than one significant peak along the line of sight to the depth of
our survey. About 120 peaks were found, ranging from rich clusters
to small groups. Here we present radial velocities and velocity dispersions
for the systems 
with at least six measured velocities.

The member galaxies were selected inside a projected circle of 
1.5 Mpc radius 
(H$_0$ = 75 km s$^{-1}$ Mpc$^{-1}$ hereafter) 
of the system center. 
For most of the systems, the center is defined as being coincident
with the position of 
the brightest cluster galaxies (BCG: a cD galaxy, a giant elliptical
or a central group of early-type galaxies), since these massive objects 
tend to be preferentially located close to the bottom of the potential well,
as an effect of dynamical friction \citep{Bir94}.
For systems without apparent dominant galaxies, the mean position  of the 
members was used.
Comparing the centers derived from the BCGs with those measured from
the overall distribution of galaxies, and with centers measured from
X-rays we find that the uncertainty of the centers is estimated to be
$\sim$ 0.1 Mpc. This comparison was done for 12 clusters,
namely A2521, A2534, A2536, A2540, A2550, A2554, A2555, A2556, A2566,
A2577, A2580 and A2606.

In the cases of galaxies with more than one measurement, these were
averaged, unless the redshift quality was poor, in which case only the
best measurement was used. The redshift extent of the individual
groups and clusters was defined using
the gaps method \citep[e.g.,][]{Zab90,Kat96}, initially selecting 
galaxies whose separation from their neighboring members was less than 
1500 km s$^{-1}$. 
This value was chosen based on the characteristic projected 
density of the sample. 
Smaller gaps, weighted by their location with respect to the middle 
of the cluster velocities distribution, 
were calculated during a second step of robust estimation 
of parameters and used iteratively to exclude remaining interlopers.

The final systemic velocity and velocity dispersion were calculated through
robust estimation of the biweighted center and scale \citep{Bee90}.
The results for 74 Aquarius clusters are presented in Table 6.
Column (1) lists the AqrCC number of the observed candidate, while 
column (2) shows another identification. Equatorial coordinates of
system center are shown in columns (3) and (4). Column (5)
indicates if the chosen projected center was the BCG(s) position or
the geometric mean of members positions.
The number of galaxies identified as members from the Aquarius Galaxy 
Catalog is shown in column (6), while the number of such 
redshifts from the literature is in column (7).
Column (8) gives the reference for literature redshifts.
The total number of member galaxies whose velocities 
were used to calculate the systemic velocity is in column (9).
Note that, since there are repeated observations from both this work
and from the literature, the total number of members with redshifts 
can be smaller than the sum of the numbers from the two sources.
Columns (10) and (11) show, respectively, the system velocity and 
velocity dispersion, both with 68\% bootstrap confidence level 
uncertainty limits.

Comparing the current data with the one on Paper I, for 25 cluster
the estimates remained unchanged, other 30 had their systemic velocities
upgraded due to inclusion of 2dFGRS velocities or more robust calculations, 
and 3 new ones were added. Major differences are presented by A2538 and
A3985 that were splitted in redshift space and APM895 and APM894 which are now 
considered a single system. Another improvement was the estimation of
velocity dispersions for the clusters.
Concerning to the work of \citet{Smi04}, they have 12 clusters in common
with Table 6 ones, most of which with a larger number of galaxies 
observed by they. 
From these, 11 have similar redshifts to the ones here. The only one with
a slightly discrepant value is A2540, for which we have measured galaxies only in 
the core of the cluster, and their value may be more representative.
Comparing the velocity dispersions of the two sources one finds that 
11 of the 12 clusters in common have higher values in their estimation.
This may also be due to the sampling of larger areas and to a less 
restrictive criteria for defining cluster members they used.
So, two effects may be playing role on this difference: our data 
underestimate whole system velocity dispersions and/or their data 
may be overestimated by residual interlopers. 
The cluster A2554, on the other hand, has higher velocity dispersion in 
our data by 188 km s$^{-1}$. Since \citet{Smi04} used almost two times
the number of member galaxies we used, and a larger area, this could 
indicate that this cluster has a declining velocity dispersion profile
on the outskirts (not common but possible). The velocity dispersion
profile of A2554 has been calculated by \citet{Fad96}, and do not
show evidence of a declining profile. The overall velocity dispersion
calculated by these authors is similar to our value 
(840$^{+131}_{-68}$ km s$^{-1}$, for 28 member galaxies). 

Finally, one should note that two clusters in our sample, 
A2540 and A2547, have estimated velocity dispersions much smaller than 
their uncertainty limits. For such clusters all the observed member 
galaxies (9 and 15, respectively) have close radial velocities, which
low probability of happen gave the high estimated uncertainty limits.
Such velocity dispersion must be used with care.
If we consider the velocity dispersion differences (excluding 
A2540 and A2554) between our and \citet{Smi04} data, we find a mean difference
of 215 km s$^{-1}$, that may be considered as the maximum 
underestimation for the whole cluster velocity dispersions of
our measurements restricted to a few more than the cluster cores.   

\section{Summary}

We presented photometry and spectroscopic redshifts for a sample
of 920 galaxies in the Aquarius region, most of which selected as
candidate members of potential groups and clusters in the region.

We described the target selection and observational procedures 
as well the redshifts estimation and error analysis. 
The results are presented in two
separated tables, one is the Aquarius Galaxy (radial velocity and 
photometric) Catalog and the other
is the cluster systemic velocity and velocity dispersion list.

The strategy of priorizing red galaxies was successful since 
85\% of the observed galaxies provided a cross-correlation 
redshift (92\% of the final catalog), and 71\% of the galaxies
were confirmed as cluster or group members.

The two main superclusters in the region, at $z \sim$ 0.086 and 
$z \sim$ 0.112, are clearly seen in the distribution of 
galaxy redshifts.

From the 81 candidates of AqrCC with observations up to date, 
we selected 58, that present clusters with at least six members
with redshifts (a total of 74 clusters due to superposition along
the line of sight), to have their systemic velocities and velocity
dispersions calculated.
The other 23 observed candidates are 14 with less than six members 
with redshifts (but indicated by their member galaxies in the Aquarius 
Galaxy Catalog), 4 that were observed but not confirmed as clusters
or groups, and 5 that appear in \citet{Smi04}.

The systemic velocities presented here are updates to the cluster 
catalog of Paper I, with the addition of cluster velocity dispersions,
but essentially the conclusions reached in that paper remain unchanged.


\acknowledgments

We are grateful to the staff and night assistants of OPD/LNA 1.6m, 
ESO 1.52m and CASLEO 2.15m; the AAO and NRL/ROE for providing the 
SSC. We also thank the referee for his helpful suggestions.
This research has also made use of NASA/IPAC Extragalactic 
Database (NED). 
C.A.C. acknowledges CNPq grant 380577/02-0, 
M.A.G.M. CNPq grants 301366/86-1, 471022/03-9 and 305529/03-0.
C.N.A.W. has been supported by the NSF grants AST 95-29098 and AST 00-71198. 



\newpage

\begin{deluxetable}{rccccrrrcrr}
\dummytable\label{tab1}
\tabletypesize{\footnotesize}
\tablewidth{0pt}
\tablenum{1}
\tablecolumns{11}
\tablecaption{Observing Runs}
\tablehead{
Run & Telescope\tablenotemark{\ast} & Date & Grating & Slit &
 Detector\tablenotemark{\dagger} & Disp. & Resol. &
 Range & Targets\tablenotemark{\ddagger} & Objects\tablenotemark{\ddagger} \\
& & & (l/mm) & ($\mu$m) & \# & (\AA/pix) & (\AA) & $\lambda\lambda$ & & \\
(1) & (2) & (3) & (4) & (5) & (6) & (7) & (8) & (9) & (10) & (11)
}
\startdata
 1 & C215 & sep94 &  316 & 500 &  Tek & 3.2 & 18 & 4760-7980 &   9 &   9 \\
 2 & O160 & oct94 &  300 & 600 &  048 & 4.1 & 14 & 4200-8260 &   9 &  11 \\
 3 & O160 & aug95 &  300 & 450 &  048 & 4.1 & 12 & 4360-8420 &  35 &  39 \\
 4 & O160 & aug96 & 600/300 & 500/400 & 101 & 4.3 & 11 & 3560-8120 & 18 &  19 \\
 5 & C215 & sep96 &  300 & 400 &  Tek & 3.2 & 16 & 4200-7660 &  35 &  38 \\
 6 & E152 & nov96 &  600 & 419 &  039 & 1.9 &  4 & 3972-7838 &  37 &  42 \\
 7 & E152 & jun97 &  600 & 419 &  039 & 1.9 &  4 & 3800-7700 &  15 &  16 \\
 8 & O160 & aug97 &  300 & 400 &  101 & 4.3 & 11 & 3650-8050 &  13 &  14 \\
 9 & E152 & aug97 &  600 & 419 &  039 & 1.9 &  4 & 3600-7500 &  53 & 117 \\
10 & O160 & oct97 &  300 & 400 &  101 & 4.3 & 11 & 3700-8100 &  17 &  25 \\
11 & E152 & oct97 &  600 & 400 &  039 & 1.9 &  4 & 3639-7537 &  12 &  29 \\
12 & E152 & jun98 &  600 & 400 &  039 & 1.9 &  4 & 3634-7518 &   2 &   6 \\
13 & E152 & aug98 &  600 & 400 &  039 & 1.9 &  4 & 3640-7520 &  52 & 121 \\
14 & O160 & aug98 &  300 & 400 &  106 & 4.3 & 11 & 3700-8100 &  23 &  36 \\
15 & E152 & oct98 &  600 & 400 &  039 & 1.9 &  4 & 3515-7514 &   4 &   9 \\
16 & O160 & aug99 &  300 & 350 &  106 & 4.3 & 10 & 3700-8100 &  19 &  36 \\
17 & E152 & aug99 &  600 & 400 &  039 & 1.9 &  4 & 3680-7560 &   2 &   5 \\
18 & O160 & sep99 &  300 & 350 &  106 & 4.3 & 10 & 3750-8150 &  28 &  46 \\
19 & E152 & oct99 & 600 & 400/350 & 038 & 1.9 & 4 & 3300-8650 & 16 &  33 \\
20 & E152 & dec99 &  600 & 400 &  038 & 1.9 &  4 & 3400-8500 &  36 &  80 \\
21 & O160 & aug00 &  300 & 250 &  106 & 4.3 &  8 & 3700-8100 &  36 &  67 \\
22 & E152 & sep00 &  600 & 400 &  038 & 1.9 &  4 & 3370-8492 &  34 &  68 \\
23 & O160 & sep00 &  300 & 250 &  106 & 4.3 &  8 & 3700-8100 &   9 &  19 \\
24 & O160 & oct00 &  300 & 250 &  106 & 4.3 &  8 & 3700-8100 &   6 &  15 \\
25 & E152 & nov00 &  600 & 250 &  038 & 1.9 &  3 & 3350-8450 &  43 &  93 \\
\enddata
\tablenotetext{\ast}{ Instrumental set up: C215 = CASLEO 2.15m + REOSC; 
O160 = OPD 1.60m + B\&C; and E152 = ESO 1.52m + B\&C}
\tablenotetext{\dagger}{ Details on table 2 }
\tablenotetext{\ddagger}{ Number of pointed ``targets'' and final number of observed
``objects'' with successful redshift measurement (due to multiobjects on the slit). }
\end{deluxetable}


\begin{deluxetable}{llccccccc}
\dummytable\label{tab2}
\tabletypesize{\footnotesize}
\tablewidth{0pt}
\tablenum{2}
\tablecolumns{9}
\tablecaption{Characteristics of the Detectors and Instrumental Set up Used}
\tablehead{
CCD & Origin & Dimensions & Pixel & Gain & Noise & Scale &
 Efficiency & Telescope \\
Nr. & & (pixels) & ($\mu$m) & ($e^{-}/$ADU) & ($e^{-}$ {\it rms}) &
 & (4000-8000) & \\
(1) & (2) & (3) & (4) & (5) & (6) & (7) & (8) & (9)
}
\startdata
      & Tek   & 1024$\times$1024 & 24   & 2.0 & 7.4 & 3.96''/100$\mu$m  & 95-70\%
& C215 \\
 \#48 & EEV   &  770$\times$1152 & 22.5 & 3.3 & 6.6 & 1.04''/100$\mu$m & 15-45\%
& O160 \\
\#101 & SITe  & 1024$\times$1024 & 24   & 5.0 & 5.5 & 1.04''/100$\mu$m & 40-65\%
& O160 \\
\#106 & SITe  & 1024$\times$1024 & 24   & 5.0 & 4.1 & 1.25''/100$\mu$m & 55-75\% 
& O160 \\
 \#39 & Loral & 2048$\times$2048 & 15   & 1.2 & 5.0 & 0.82''/100$\mu$m & 75-95\%
& E152 \\
 \#38 & Loral & 2688$\times$512  & 15   & 1.6 & 7.1 & 0.82''/100$\mu$m & 75-90\%
& E152 \\
\enddata
\end{deluxetable}


\begin{deluxetable}{lrccc}
\dummytable\label{tab3}
\tabletypesize{\footnotesize}
\tablewidth{0pt}
\tablenum{3}
\tablecolumns{5}
\tablecaption{Average and Median Internal Errors\tablenotemark{\ast}}
\tablehead{
Telescope & N & $<\delta$$v>$ & $\sigma_v$ & [$\delta$$v$]
}
\startdata
E152     & 581 & 58 & 44 & 48 \\
O160     & 310 & 66 & 39 & 57 \\
C215     &  29 & 87 & 52 & 75 \\
all data & 920 & 62 & 43 & 51 \\
\enddata
\tablenotetext{\ast}   { $<>$ e [\,] represent, respectively, mean and median values. }
\end{deluxetable}


\begin{deluxetable}{lccrrrrccrrrrccrrrrccrrrccccrrr}
\dummytable\label{tab4}
\tabletypesize{\footnotesize}
\tablewidth{0pt}
\tablenum{4}
\tablecolumns{31}
\setlength{\tabcolsep}{0.02in}
\tablecaption{Velocity Standard Galaxies\tablenotemark{\ast}}
\tablehead{
 \multicolumn{1}{c}{Galaxy} & & \multicolumn{22}{c}{This work} & & 
\multicolumn{6}{c}{NED} \\
 \cline{3-24} \cline{26-31} \\[-0.2 cm]
 & & \multicolumn{5}{c}{E152} & & \multicolumn{5}{c}{O160} & & 
\multicolumn{5}{c}{C215} & & \multicolumn{4}{c}{all data} & & 
\multicolumn{2}{c}{Small Error}\tablenotemark{\dagger} & & \multicolumn{3}{c}{all data} \\
 \cline{3-7} \cline{9-13} \cline{15-19} \cline{21-24} \cline{26-27} \cline{29-31}\\[-0.2 cm]
 & & $<v>$ & $\sigma_v$ & N$_{r}$ & N$_{o}$ & $\Delta$$v$ & & 
$<v>$ & $\sigma_v$ & N$_{r}$ & N$_{o}$ & $\Delta$$v$ & & 
$<v>$ & $\sigma_v$ & N$_{r}$ & N$_{o}$ & $\Delta$$v$ & & 
$<v>$ & $\sigma_v$ & N$_{o}$ & $\Delta$$v$ & & 
$v$ & Ref.\tablenotemark{\S} & & $<v>$ & $\sigma_v$ & N$_{v}$ \\
 \multicolumn{1}{c}{(1)} & & (2) & (3) & (4) & (5) & (6) & & (7) & (8) & (9) & (10) & (11) & & 
(12) & (13) & (14) & (15) & (16) & & (17) & (18) & (19) & (20) & & 
(21) & (22) & & (23) & (24) & (25) \\[-0.3 cm]
}
\startdata
NGC 7507 & & 1594 & 20 &  3 & 12 &  +28 & & 1616 & 43 &  1 &  3 &  +50 & & ... & ... & ... & ... & ... & & 
 1605 & 25 & 15 &  +39 & & 1566$\pm$15 & 1 & & 1577 & 29 & 10 \\
NGC 1316 & & 1752 & 21 &  2 & 18 & $-$8 & & ... & ... & ... & ... & ... & & ... & ... & ... & ... & ... & & 
 1760 & 11 & 18 &   +0 & & 1760$\pm$10 & 2 & & 1795 & 54 & 14 \\
NGC 6958 & & 2738 & 29 &  7 & 18 &  +25 & & ... & ... & ... & ... & ... & & ... & ... & ... & ... & ... & & 
 2735 & 30 & 18 &  +22 & & 2713$\pm$13 & 1 & & 2687 & 53 &  7 \\
NGC 6868 & & 2926 & 12 &  4 & 10 &  +72 & & 2913 & 12 &  1 &  7 &  +59 & & ... & ... & ... & ... & ... & & 
 2917 & 43 & 17 &  +63 & & 2854$\pm$15 & 3 & & 2843 & 47 &  8 \\
NGC 5419 & & ... & ... & ... & ... & ... & & 4196 & 62 & 3 & 12 &  +70 & & ... & ... & ... & ... & ... & & 
 4173 & 57 & 12 &  +47 & & 4126$\pm$15 & 4 & & 4136 & 77 &  7 \\
NGC 6721 & & 4435 & 13 &  2 & 11 &  +14 & & 4486 & 44 &  7 & 43 &  +65 & & 4390 & 28 &  1 &  7 & -31 & &  
 4455 & 59 & 61 &  +34 & & 4421$\pm$25 & 1 & & 4421 & 47 &  5 \\
NGC 6841 & & 5845 & 23 &  3 &  6 & $-$41 & & 5798 & 9 &  1 &  4 & $-$88 & & ... & ... & ... & ... & ... & & 
 5828 & 30 & 10 & $-$58 & &       5886 & 5 & & ... & ... &  1 \\ 
NGC 0641 & & ... & ... & ... & ... & ... & & 6437 & 28 & 4 & 9 &  +114 & & ... & ... & ... & ... & ... & & 
 6438 & 29 &  9 & +115 & & 6323$\pm$17 & 1 & & 6328 & 59 &  7 \\
\cline{7-7} \cline{13-13} \cline{19-19} \cline{24-24} \\[-0.2 cm]
Averages & & & & & & +15 & & & & & & +45 & & & & & & -31 & & & & & +33 & & & & & & &  
\enddata
\tablenotetext{\ast}   { $<v>$ represents mean measured velocity; $\sigma_v$, its standard deviation;
 N$_{r}$, the number of runs in which the galaxy was observed; N$_{o}$, the number of observations; 
 N$_{v}$, the number of velocities in the NED database; and $\Delta$$v$, the difference between the
 mean velocity and the respective NED ``small error'' velocity. }
\tablenotetext{\dagger}{ Velocity on NED with the smallest error. }
\tablenotetext{\S}     { Reference to the smallest error NED velocity:
(1) {\it Third Reference Catalogue of Bright Galaxies} (RC3), 1991 (version 3.9);
(2) \citet{Lon98};
(3) \citet{Ram96};
(4) \citet{Kal03}:
(5) \citet{Alo03}. }
\end{deluxetable}


\begin{deluxetable}{rccrrlcrlr}
\dummytable\label{tab5}
\tabletypesize{\footnotesize}
\tablewidth{0pt}
\tablenum{5}
\tablecolumns{10}
\tablecaption{Aquarius Galaxy Catalog}
\tablehead{
Gal & \multicolumn{2}{c} {Coordinates (J2000.0)} & $b_J$ & $R$ &
 Cluster & c$z$ & Error & Code\tablenotemark{\ast} & Run\tablenotemark{\dagger} \\
\cline{2-3} \\[-0.3cm]
Nr. & \colhead{$\alpha$} &
 \colhead{$\delta$} &
 mag. & mag. & name & ($km s^{-1}$) & ($km s^{-1}$) & & \\
(1) & (2) & (3) & (4) & (5) & (6) & (7) & (8) & (9) & (10)
}
\startdata
  1 &  22$^h$57$^m$53.61$^s$ & -21$\degr$46$\arcmin$57.6$\arcsec$ & 18.44 & \nodata & A2509B  &   23915 &   55
 & x,A, 6.21 & 21 \\
  2 &  22 57 56.77 & -21 50 01.9 &   18.09 & \nodata & A2509B  &   25503 &   56 & c,A, 2.89 & 21 \\
  3 &  22 57 57.54 & -21 46 37.9 &   19.12 & \nodata & A2509B  &   23762 &   70 & e,A, 0.00 & 21 \\
  4 &  22 58 02.21 & -21 53 58.2 &   18.65 & \nodata & A2509B  &   24007 &   71 & c,A, 4.00 & 12 \\
  5 &  22 58 02.58 & -21 53 51.1 &   18.62 & \nodata & A2509C  &   40169 &   19 & x,A,14.59 & 12 \\
  6 &  22 58 02.96 & -21 54 01.4 &   19.62 & \nodata & A2509C  &   40132 &   26 & x,A,10.44 & 12 \\
  7 &  22 58 03.09 & -21 53 45.1 &   19.80 & \nodata & A2509C  &   40253 &   56 & c,B, 4.18 & 12 \\
  8 &  22 58 05.22 & -21 54 29.5 &   19.55 & \nodata & A2509A  &   69163 &   34 & x,A, 8.00 & 12 \\
  9 &  22 58 06.56 & -21 54 45.6 &   19.36 & \nodata & A2509A  &   69774 &   31 & x,A,10.78 & 12 \\
 10 &  22 58 07.41 & -21 48 34.1 &   18.51 & \nodata & A2509B  &   23640 &  129 & c,B, 2.39 & 21 \\
 11 &  22 58 08.06 & -21 55 02.7 & \nodata & \nodata & \nodata &   63541 &   77 & e,B, 2.46 & 12 \\
 12 &  22 58 15.34 & -21 56 19.6 & \nodata & \nodata & A2509B  &   23787 &   63 & c,A, 3.31 & 21 \\
 13 &  22 58 16.57 & -23 33 44.9 &   18.48 & \nodata & Aqr005  &   35160 &   34 & x,A, 9.72 & 22 \\
 14 &  22 58 20.84 & -23 33 38.5 &   19.24 & \nodata & Aqr005  &   35449 &   38 & x,A, 8.57 & 22 \\
 15 &  22 58 21.16 & -23 36 47.6 &   17.83 & \nodata & Aqr005  &   34775 &   39 & c,A, 7.00 & 22 \\
 16 &  22 58 21.34 & -21 52 48.8 &   19.81 & \nodata & \nodata &   55460 &   63 & x,A, 4.23 & 21 \\
 17 &  22 58 26.71 & -23 37 57.4 &   18.82 & \nodata & Aqr005  &   34847 &   44 & x,A, 7.65 & 22 \\
 18 &  22 58 27.06 & -21 54 18.2 &   18.85 & \nodata & A2509C  &   40372 &   57 & c,A, 3.98 & 21 \\
 19 &  22 58 30.50 & -23 39 09.4 &   18.03 & \nodata & Aqr005  &   35541 &   43 & c,A, 6.47 & 22 \\
 20 &  22 58 30.62 & -23 42 33.2 &   18.72 & \nodata & Aqr005  &   35100 &   33 & x,A,11.31 & 22 \\
\enddata
\tablenotetext{\ast}{ Codes are for: type of spectrum (x = cross-correlation, e = emission lines,
c = combined absorption and emission lines); quality flag (A,B or C); and ``R'' parameter. }
\tablenotetext{\dagger}{ Refers to Table 1. }
\tablecomments{The complete version of this table is in the electronic edition of the journal.
The printed edition contains only a sample. }
\end{deluxetable}


\begin{deluxetable}{clccccrrrrcc}
\dummytable\label{tab6}
\tabletypesize{\footnotesize}
\tablewidth{0pt}
\tablenum{6}
\tablecolumns{12}
\tablecaption{Cluster Data}
\tablehead{
{AqrCC} & {Other} & \multicolumn{3}{c}{Coordinates (J2000.0)} & &
   \multicolumn{4}{c}{Observed Galaxies}          & {$v_{r}$} & {$\sigma_{r}$} \\
\cline{3-5} \cline{7-10} \\[-0.25cm]
{}    & {Name}  &   $\alpha$   &   $\delta$   & Center\tablenotemark{\ast} & &
{$N_{here}$} & {$N_{lit}$} & {Ref.\tablenotemark{\dagger}} & {$N_{gal}$} & (km s$^{-1}$) & (km s$^{-1}$) \\
(1)   & (2)     & (3)          & (4)          & (5)         & &
(6)          & (7)         & (8)    & (9)        & (10)        & (11)
}
\startdata
001 & Aqr001-A & 22$^{h}$57$^{m}$14.6$^{s}$ & -24$^{\circ}$58$'$20$'\!'$ & G & & ... & 7 & 13,16 & 7 &  26542$^{+  64}_{ -82}$ &    184$^{+  83}_{ -15}$ \\
001 & Aqr001-B  & 22 58 16.7 & -25 00 56 & G   & & ... &  12 &       16 & 12 &  27348$^{+  50}_{ -18}$ &    114$^{+  43}_{ -23}$ \\
005 & Aqr005    & 22 58 32.0 & -23 37 37 & B?  & &  11 &   7 &       16 & 15 &  35406$^{+  86}_{-129}$ &    393$^{+  62}_{ -40}$ \\
004 & A3949     & 22 58 52.1 & -19 57 17 & G   & &   7 & ... &      ... &  7 &  47358$^{+  60}_{-210}$ &    207$^{+1008}_{ -30}$ \\
009 & A2518-D   & 23 00 31.9 & -24 19 41 & G   & & ... &   6 &       16 &  6 &  35143$^{+  90}_{-292}$ &    383$^{+ 139}_{ -67}$ \\
009 & A2518-C   & 23 00 44.7 & -24 06 22 & G   & & ... &   7 &       16 &  7 &  32953$^{+  74}_{  -3}$ &     57$^{+  85}_{ -37}$ \\
009 & A2518     & 23 00 47.0 & -24 09 02 & B2  & &   6 &  15 &     3,16 & 17 &  40471$^{+ 126}_{-149}$ &    547$^{+ 158}_{ -90}$ \\
009 & A2518-B   & 23 00 57.8 & -24 12 08 & G   & &   4 &   5 &       16 &  7 &  27595$^{+  42}_{-172}$ &    199$^{+ 645}_{-118}$ \\
010 & A2521     & 23 02 11.4 & -22 01 24 & B2  & &  14 &   5 &    2,3,5 & 17 &  40883$^{+ 463}_{-504}$ &    248$^{+ 289}_{ -89}$ \\
012 & A2526     & 23 04 34.9 & -24 04 11 & B2  & &   6 &   4 &       16 &  8 &  60985$^{+  69}_{-102}$ &    376$^{+ 332}_{-206}$ \\
014 & A2527     & 23 05 20.6 & -25 20 19 & B2  & &   8 &   4 &       16 & 10 &  48691$^{+ 207}_{-585}$ &    648$^{+ 153}_{ -40}$ \\
015 & A2528     & 23 05 36.0 & -21 23 03 & B   & &  11 &   1 &        3 & 12 &  28475$^{+ 165}_{-197}$ &    458$^{+  85}_{ -44}$ \\
016 & A2531     & 23 06 56.0 & -21 39 47 & B   & &   6 &   1 &        3 &  7 &  51873$^{+ 343}_{-198}$ &    560$^{+ 236}_{ -59}$ \\
017 & A2534     & 23 07 41.5 & -22 42 39 & B   & &  15 &   1 &        3 & 16 &  60348$^{+ 328}_{-238}$ &    126$^{+ 479}_{-263}$ \\
019 & Aqr019-C  & 23 07 42.4 & -25 27 12 & G   & & ... &   8 &       16 &  8 &  34635$^{+ 339}_{-242}$ &    571$^{+ 175}_{ -30}$ \\
018 & A2536     & 23 07 46.7 & -22 27 31 & B   & &   7 &   1 &        3 &  6 &  59811$^{+1143}_{-511}$ &    981$^{+ 206}_{ -18}$ \\
018 & A2536-B   & 23 07 49.9 & -22 29 09 & G   & &   9 & ... &      ... &  8 &  52575$^{+ 132}_{-181}$ &    353$^{+ 150}_{ -74}$ \\
019 & Aqr019-A  & 23 08 03.1 & -25 41 01 & G   & & ... &   9 &       16 &  9 &  30860$^{+  45}_{-110}$ &    313$^{+ 365}_{-170}$ \\
020 & A2538-2   & 23 08 28.5 & -19 51 20 & G   & & ... &  23 &    3,4,9 & 23 &  25512$^{+  91}_{ -77}$ &    374$^{+  74}_{ -57}$ \\
020 & A2538-1   & 23 08 34.1 & -19 52 34 & B2  & & ... &  21 &      4,9 & 21 &  24099$^{+ 136}_{-137}$ &    465$^{+  78}_{ -55}$ \\
019 & Aqr019-B  & 23 08 42.3 & -25 30 47 & G   & & ... &   6 &       16 &  6 &  33017$^{+  94}_{-181}$ &    274$^{+ 135}_{ -55}$ \\
022 & Aqr022    & 23 09 03.6 & -20 45 15 & G   & &   5 &   1 &       15 &  6 &  24739$^{+  69}_{ -38}$ &    152$^{+  61}_{ -34}$ \\
023 & A2540     & 23 09 25.1 & -22 10 19 & B   & &   8 &   1 &        3 &  9 &  38875$^{+  21}_{ -16}$ &     38$^{+ 555}_{ -12}$ \\
024 & Aqr024    & 23 09 54.5 & -21 31 30 & B   & &   7 & ... &      ... &  7 &  33250$^{+ 275}_{-280}$ &    600$^{+ 210}_{ -68}$ \\
026 & A2542     & 23 10 06.2 & -24 29 09 & B2  & &   4 &   6 &       16 &  6 &  50560$^{+ 287}_{  -8}$ &    454$^{+ 234}_{-439}$ \\
027 & Aqr027-C  & 23 10 06.9 & -24 44 19 & G   & &   1 &   6 &       16 &  6 &  36060$^{+ 233}_{-445}$ &    715$^{+ 238}_{ -84}$ \\
025 & A2541     & 23 10 07.2 & -22 59 05 & B2  & &  17 &   2 &        8 & 17 &  34032$^{+ 243}_{-284}$ &    991$^{+ 138}_{ -87}$ \\
027 & Aqr027-B  & 23 10 13.1 & -24 46 35 & G   & &   2 &   9 &       16 & 10 &  32625$^{+ 526}_{-375}$ &    785$^{+ 188}_{ -69}$ \\
028 & A2546     & 23 10 38.7 & -22 38 54 & B2  & &  22 &   1 &        3 & 22 &  33903$^{+ 223}_{-177}$ &    784$^{+ 164}_{ -98}$ \\
027 & Aqr027    & 23 10 39.9 & -24 45 13 & B2  & &   5 &   3 &       16 &  6 &  59786$^{+ 195}_{-252}$ &    465$^{+ 110}_{ -46}$ \\
029 & A2547     & 23 10 46.7 & -21 08 07 & B2  & &  12 &   5 &  3,12,15 & 15 &  45483$^{+ 409}_{-476}$ &     80$^{+ 272}_{-136}$ \\
031 & A2548     & 23 11 15.8 & -20 25 05 & B   & &   6 &   4 &     3,15 &  9 &  33105$^{+  81}_{ -46}$ &    180$^{+  73}_{ -15}$ \\
033 & A2550     & 23 11 35.8 & -21 44 47 & B   & &   5 &   2 &     3,15 &  6 &  36752$^{+ 321}_{-329}$ &    517$^{+ 230}_{ -22}$ \\
035 & A2554     & 23 12 19.9 & -21 30 10 & B   & &   4 &  31 &   3,4,15 & 32 &  33243$^{+ 139}_{-166}$ &    868$^{+ 129}_{ -68}$ \\
036 & A2553     & 23 12 24.8 & -24 57 12 & B2  & & ... &  13 &    10,16 & 13 &  44081$^{+ 191}_{-198}$ &    686$^{+ 161}_{ -76}$ \\
037 & A2555     & 23 12 51.3 & -22 15 27 & B?  & &   9 &   2 &       15 & 11 &  33236$^{+  25}_{ -85}$ &    238$^{+  84}_{ -72}$ \\
038 & A2556     & 23 13 01.5 & -21 38 04 & B   & &   5 &   8 & 1,3,4,15 & 10 &  26109$^{+  75}_{-260}$ &    352$^{+ 355}_{-112}$ \\
039 & S1099     & 23 13 10.9 & -23 09 29 & B2  & &  12 & ... &      ... & 12 &  33258$^{+ 373}_{-227}$ &    732$^{+ 238}_{ -98}$ \\
040 & Aqr040    & 23 14 30.8 & -23 22 24 & B2  & &   6 &   1 &       16 &  7 &  27274$^{+  53}_{ -42}$ &    122$^{+  47}_{ -20}$ \\
041 & A2565-A   & 23 15 52.0 & -21 08 30 & G   & &  12 & ... &      ... & 12 &  24727$^{+  69}_{-144}$ &    359$^{+ 483}_{ -43}$ \\
041 & A2565-B   & 23 15 54.9 & -21 08 18 & G   & &  10 &   1 &        3 & 10 &  38365$^{+  86}_{-169}$ &    325$^{+  76}_{ -51}$ \\
044 & A2566     & 23 16 05.0 & -20 27 48 & B   & &  10 &   2 &      3,9 & 11 &  24642$^{+ 159}_{-439}$ &    856$^{+ 179}_{-154}$ \\
042 & A3985-2   & 23 16 12.1 & -23 22 55 & G   & &   8 &   2 &       16 & 10 &  33940$^{+  64}_{ -74}$ &    181$^{+ 233}_{ -34}$ \\
042 & A3985-1   & 23 16 15.1 & -23 23 36 & B   & &  10 &   5 &       16 & 12 &  31855$^{+ 189}_{-139}$ &    482$^{+ 115}_{ -63}$ \\
045 & Aqr045-A  & 23 16 20.9 & -25 00 29 & G   & & ... &   9 &       16 &  9 &  15436$^{+  21}_{-356}$ &    329$^{+ 104}_{ -41}$ \\
045 & Aqr045-B  & 23 16 32.1 & -24 57 14 & G   & & ... &   7 &       16 &  7 &  33000$^{+ 581}_{-235}$ &    513$^{+ 190}_{ -58}$ \\
046 & A2568     & 23 17 11.8 & -22 14 23 & B?  & &   5 &   1 &        3 &  6 &  41875$^{+ 102}_{-232}$ &    327$^{+ 280}_{-155}$ \\
047 & ED275     & 23 17 33.7 & -25 20 30 & B2  & &   6 &   9 &       16 & 12 &  43329$^{+ 241}_{-293}$ &    670$^{+ 192}_{ -60}$ \\
051 & A2576     & 23 19 43.9 & -22 27 52 & B?  & & ... &  10 &        7 & 10 &  56253$^{+ 453}_{-458}$ &    750$^{+ 193}_{ -65}$ \\
052 & S1113     & 23 20 01.8 & -24 08 46 & B?  & &   9 &  10 &       16 & 14 &  44005$^{+ 118}_{ -98}$ &    367$^{+  53}_{ -47}$ \\
055 & A2577     & 23 20 46.7 & -22 59 30 & B   & &   6 &   1 &       10 &  7 &  37363$^{+  61}_{-278}$ &    176$^{+  92}_{  -9}$ \\
057 & A2579     & 23 21 15.6 & -21 35 04 & B   & &   4 &   6 &    10,15 & 10 &  33409$^{+ 104}_{-143}$ &    394$^{+  92}_{ -46}$ \\
061 & A2580     & 23 21 26.3 & -23 12 26 & B   & &  19 & ... &      ... & 17 &  26687$^{+ 173}_{-164}$ &    661$^{+ 130}_{ -82}$ \\
059 & A3997     & 23 21 33.7 & -24 08 52 & B3  & &   8 &   5 &       16 & 11 &  44280$^{+ 133}_{ -90}$ &    391$^{+ 455}_{-166}$ \\
063 & A2583     & 23 22 14.7 & -20 26 08 & B?  & &   8 & ... &      ... &  8 &  34388$^{+ 237}_{-293}$ &    659$^{+ 264}_{-104}$ \\
065 & A2586     & 23 23 24.3 & -20 22 35 & B   & &  11 & ... &      ... & 11 &  43420$^{+ 316}_{ -77}$ &    828$^{+ 274}_{-261}$ \\
066 & A2587     & 23 23 32.4 & -22 25 21 & B   & &   6 & ... &      ... &  6 &  64750$^{+ 807}_{-131}$ &    673$^{+ 287}_{-101}$ \\
067 & ED291     & 23 24 13.2 & -22 31 12 & G   & &  15 &   1 &       15 & 15 &  36690$^{+  94}_{-127}$ &    378$^{+  88}_{ -43}$ \\
070 & A2596     & 23 25 05.6 & -23 23 49 & B2  & &  21 &  10 & 13,15,16 & 27 &  26719$^{+ 113}_{-117}$ &    496$^{+  79}_{ -36}$ \\
069 & A2595     & 23 25 06.2 & -20 32 37 & B?  & &   5 & ... &      ... &  5 &  54091$^{+ 199}_{-561}$ &    545$^{+ 173}_{ -21}$ \\
073 & APM895/894 & 23 26 14.1 & -24 06 30 & B & & 16 & 36 & 11,14,15,16 & 35 &  33463$^{+  67}_{-102}$ &    549$^{+ 109}_{ -72}$ \\
075 & A2600     & 23 26 31.9 & -22 25 47 & G   & &   6 &   3 &       15 &  9 &  36268$^{+ 344}_{-164}$ &    570$^{+ 410}_{-125}$ \\
074 & A2599-B   & 23 26 38.1 & -23 46 04 & B   & &   9 &   9 & 14,15,16 & 13 &  37841$^{+ 109}_{-226}$ &    585$^{+ 344}_{-111}$ \\
074 & A2599-A   & 23 26 54.5 & -23 51 46 & G   & & 9 & 26 & 6,8,14,15,16 & 22 & 26943$^{+ 128}_{ -86}$ &    456$^{+ 105}_{ -63}$ \\
076 & A2601     & 23 27 01.8 & -24 30 04 & G   & &   4 &   4 &    15,16 &  7 &  63552$^{+  72}_{-307}$ &    635$^{+ 597}_{-219}$ \\
075 & A2600-B   & 23 27 11.2 & -22 18 55 & G   & &   2 &   7 &       15 &  9 &  18305$^{+ 261}_{-167}$ &    363$^{+  91}_{ -84}$ \\
078 & ED300-B   & 23 27 45.4 & -25 05 42 & G   & &   1 &  11 &       16 & 11 &  26436$^{+ 170}_{-521}$ &    449$^{+ 139}_{ -68}$ \\
077 & A2603     & 23 27 56.8 & -25 22 37 & B   & &   7 &   3 &      ... &  6 &  63218$^{+ 110}_{-115}$ &    396$^{+ 174}_{-144}$ \\
078 & ED300     & 23 28 16.2 & -24 56 43 & B   & &  12 &  12 &       16 & 18 &  33480$^{+ 173}_{-143}$ &    585$^{+  94}_{ -41}$ \\
081 & A2605-A   & 23 28 48.9 & -23 22 22 & B2  & &  11 &   1 &       15 & 10 &  33589$^{+ 222}_{-222}$ &    545$^{+ 154}_{ -58}$ \\
081 & A2605-B   & 23 29 05.9 & -23 25 08 & B   & &   6 & ... &      ... &  6 &  26621$^{+ 166}_{-170}$ &    266$^{+ 108}_{ -22}$ \\
091 & A4014     & 23 32 25.7 & -25 28 50 & B2  & &  11 &  16 &       16 & 22 &  33840$^{+  57}_{ -46}$ &    261$^{+ 132}_{ -37}$ \\
099 & A2628     & 23 37 01.4 & -24 10 58 & G   & & ... &   9 &        7 &  9 &  55468$^{+ 217}_{-448}$ &    810$^{+ 247}_{-121}$ \\
101 & A2629     & 23 37 47.2 & -22 54 42 & B?  & &   7 & ... &      ... &  7 &  62038$^{+ 405}_{-470}$ &    986$^{+ 327}_{ -68}$ \\
\enddata
\tablenotetext{\ast}{ Type of projected center: (B\#) position of the brightest cluster galaxy (\# is the
number of giant ellipticals considered as BCGs of the cluster, ? means BCG is not surely defined); 
(G) geometric mean of the positions of 
galaxies confirmed as cluster members. }
\tablenotetext{\dagger}{ References of the galaxy velocities from the literature: (1) \citet{KUC83}; 
(2) \citet{SGH83}; (3) \citet{CFH85}; (4) \citet{CeH87}; (5) \citet{VeC88}; (6) \citet{Dal94};
(7) \citet{Bat95}; (8) \citet{Col95}; (9) \citet{OLK95}; (10) \citet{QeR95}; (11) \citet{Lov96};
(12) \citet{Kap98}; (13) \citet{Rat98}; (14) \citet{Alo99}; (15) \citet{Bat99}; (16) \citet{2dF}. }
\end{deluxetable}


\end{document}